# On cycles through two arcs in strong multipartite tournaments

Alexandru I. Tomescu[*]

**Abstract**

A multipartite tournament is an orientation of a complete $c$-partite graph. In [L. Volkmann, A remark on cycles through an arc in strongly connected multipartite tournaments, Appl. Math. Lett. 20 (2007) 1148–1150], Volkmann proved that a strongly connected $c$-partite tournament with $c \geqslant 3$ contains an arc that belongs to a directed cycle of length $m$ for every $m \in \{3, 4, \ldots, c\}$. He also conjectured the existence of three arcs with this property. In this note, we prove the existence of two such arcs.

**Keywords:** multipartite tournament; cycle; cycle through an arc.

A *c-partite* or *multipartite tournament* is an orientation of a complete $c$-partite graph. By a *cycle* or *path* we mean a simple directed cycle or simple directed path. For standard terminology on directed graphs see, e.g., Bang-Jensen and Gutin [1]. In what follows, all digraphs are finite, without loops or multiple arcs. The vertex set of a digraph $D$ is denoted by $V(D)$. If $xy$ is an arc of a digraph $D$, then we write $x \to y$ and say that $x$ dominates $y$. If $X$ and $Y$ are two disjoint subsets of $V(D)$, such that every vertex of $X$ dominates every vertex of $Y$, then we say that $X$ dominates $Y$, and write $X \to Y$. A digraph $D$ is *strongly connected* or *strong* if for each pair of vertices $u$ and $v$ there is a path in $D$ from $u$ to $v$. A cycle of length $m$ is also called an $m$-cycle. A digraph $D$ is *pancyclic* if it contains an $m$-cycle for all $m$ between 3 and $|V(D)|$. A vertex or an arc is pancyclic in a digraph $D$ if it belongs to an $m$-cycle for all $m$ between 3 and $|V(D)|$.

Moon [2] obtained the following result on pancyclic arcs for strongly connected tournaments.

**Theorem 1** *(Moon [2]) Every strong tournament contains at least three pancyclic arcs.*

In [3], Volkmann showed that a similar result holds for the case of strong multipartite tournaments.

**Theorem 2** *(Volkmann [3]) If $D$ is a strong $c$-partite tournament with $c \geqslant 3$, then $D$ contains at least one arc that belongs to an $m$-cycle $C_m$ for each $m \in \{3, 4, \ldots, c\}$ such that $V(C_3) \subset V(C_4) \subset \cdots \subset V(C_c)$.*

Volkmann also proposed [3, 4] a possible improvement of this result, by formulating the following conjecture.

---

[*]Dipartimento di Matematica e Informatica, Università di Udine, Via delle Scienze, 206, 33100 Udine, Italy, alexandru.tomescu@uniud.it

Faculty of Mathematics and Computer Science, University of Bucharest, Str. Academiei, 14, 010014 Bucharest, Romania

**Conjecture 1** *(Volkmann [3]) If $D$ is a strong c-partite tournament with $c \geqslant 3$, then $D$ contains at least three arcs that belong to an m-cycle for each $m \in \{3, 4, \ldots, c\}$.*

In this note we will prove a weaker version of this conjecture, by showing that in a $c$-partite strong tournament there are two arcs that belong to an $m$-cycle for each $m \in \{3, 4, \ldots, c\}$. Note that the proof of Theorem 3 below works along the same lines as the proof of Theorem 2.

**Theorem 3** *If $D$ is a strong c-partite tournament with $c \geqslant 3$, then $D$ contains two arcs $e^1, e^2$, where each $e^k \in \{e^1, e^2\}$ belongs to an m-cycle $C_m^k$, for each $m \in \{3, 4, \ldots, c\}$, such that $V(C_3^k) \subset V(C_4^k) \subset \cdots \subset V(C_c^k)$.*

**Proof**: It is known that $D$ contains a 3-cycle (see, for example, [3]). Any two arcs of this cycle satisfy the claim for $m = 3$.

Suppose now that for some $m$ satisfying $3 \leqslant m < c$, there are two arcs $e^1$ and $e^2$, such that each $e^k \in \{e^1, e^2\}$ belongs to $m - 2$ cycles $C_3^k, C_4^k, \ldots, C_m^k$ with: $|V(C_j^k)| = j$, for all $3 \leqslant j \leqslant m$, and $V(C_3^k) \subset V(C_4^k) \subset \cdots \subset V(C_m^k)$.

In what follows, we will show that for each $k \in \{1, 2\}$ there are two ways to satisfy the inductive requirement. First, $e^k$ can also be contained in a cycle $C_{m+1}^k$ of length $m + 1$, such that $V(C_m^k) \subset V(C_{m+1}^k)$. Second, we can find two arcs $f^1, f^2$, belonging to a $j$-cycle for each $3 \leqslant j \leqslant m + 1$, with the desired properties.

Let $e \in \{e^1, e^2\}$ and let $C_m = u_1 u_2 \ldots u_m u_1$, where $e = u_1 u_2$, be the cycle of length $m$ obtained from the inductive hypothesis. Let $S$ be the set of vertices that belong to partite sets not represented on $C_m$.

If $S$ contains a vertex $w$ that has an out-neighbor and an in-neighbor on $C_m$, then $w$ has an in-neighbor $u^-$ on $C_m$ immediately followed by an out-neighbor $u^+$ on $C_m$. If $u^- \neq u_1$ and $u^+ \neq u_2$, then we can form an $(m+1)$-cycle $C_{m+1}$ containing $e$ by replacing the arc $u^- u^+$ of $C_m$ with the arcs $u^- w, w u^+$. Observe that $V(C_m) \subset V(C_{m+1})$.

Assume now that $u^- = u_1$ and $u^+ = u_2$, and there are no further two vertices on $C_m$ with these properties. Observe first that there exists an index $2 \leqslant i \leqslant m$ such that $w \to \{u_2, u_3, \ldots, u_i\}$ and $\{u_{i+1}, u_{i+2}, \ldots, u_m, u_1\} \to w$. The arc $u_i u_{i+1}$ is contained in cycles of lengths $3, 4, \ldots, m+1$ with the desired properties, because $u_i u_{i+1} \ldots u_{i+j} w u_i$, where $u_{m+1} = u_1$ are cycles of lengths $3, 4, \ldots, m+3-i$ containing $u_i u_{i+1}$, for $1 \leqslant j \leqslant m+1-i$ and $u_{i-k} u_{i+1-k} \ldots u_i u_{i+1} \ldots u_m u_1 w u_{i-k}$ are cycles of length $m+4-i, m+5-i, \ldots, m+1$ containing $u_i u_{i+1}$ for $1 \leqslant k \leqslant i-2$, when $i \geqslant 3$.

Second, as $u_1 u_2$ belongs to a 3-cycle, there exists a vertex $v \in V(D)$ such that $u_1 u_2 v u_1$ is a cycle. From the inductive hypothesis, we also have $v \in V(C_m)$. As $v$ belongs to a different partite set than $w$, we have either $w \to v$, or $v \to w$. In the former case, the arc $u_1 w$ belongs to the 3-cycle $C_3' = u_1 w v u_1$, while in the latter case, the arc $w u_2$ belongs to the 3-cycle $C_3' = w u_2 v w$. Additionally, both $u_1 w$ and $w u_2$ are contained in the 4-cycle $C_4' = u_1 w u_2 v u_1$. In general, as $V(C_3) \subset V(C_4) \subset \cdots \subset V(C_m)$ and the fact that $w$ belongs to a partite set not represented on $C_m$, every cycle $C_i$ of length $i$, $4 \leqslant i \leqslant m$, containing $u_1 u_2$ can be extended to a cycle $C_{i+1}'$ of length $i+1$, by replacing the arc $u_1 u_2$ with the two arcs $u_1 w$ and $w u_2$. Moreover, we have $V(C_3') \subset V(C_4') \subset \cdots \subset V(C_{m+1}')$. In conclusion, we can replace the arcs $e^1, e^2$ by the arc $f^1 = u_i u_{i+1}$ and an arc $f^2 \in \{u_1 w, w u_2\}$, $f^2 \neq f^1$, and satisfy our claim.

The only case it remains to consider is when $S$ can be decomposed into two subsets $S_1$ and $S_2$ such that $S_2 \to V(C_m) \to S_1$. Since $m < c$, we may assume, without loss of generality, that $S_1 \neq \emptyset$. As $D$ is strongly connected, there is a path from $S_1$ to $C_m$. Let $P = y_1 y_2 \ldots y_q$ be a shortest such path, where $q \geqslant 3$, $y_1 \in S_1$ and $y_q \in V(C_m)$. We note that $y_t \notin S_1 \cup S_2$, for all $2 \leqslant t \leqslant q - 2$, as otherwise the minimality of $P$ would be contradicted. Therefore, we also have $y_t \to y_1$, for all $2 \leqslant t \leqslant q - 2$.

When $y_{q-1} \in S_2$, we have an arc between $y_{q-1}$ and every vertex on $C_m$. Let $u_z$ be a vertex of $C_m$ such that $u_z$ and $y_{q-2}$ are in different partite sets. This entails $u_z \to y_{q-2}$. The arcs $f^1 = u_z y_{q-2}$ and $f^2 = y_{q-2} y_{q-1}$ are both contained in cycles of length $3, 4, \ldots, m+1$ because $u_z y_{q-2} y_{q-1} u_j u_{j+1} \ldots u_{z-1} u_z$ are cycles of length $m + 3 - j$, for $2 \leqslant j \leqslant m$, with the desired property.

Consider now $y_{q-1} \notin S_2$. If $q > 3$, the minimality of $P$ implies that $y_{q-1}$ and $y_1$ belong to different partite sets, and hence $y_{q-1} \to y_1$. If $q = 3$ then $y_3 \in C_m$, and since $y_1 \in S_1$ and $V(C_m) \to S_1$, we have $y_3 \to y_1$. Therefore, the arcs $f^1 = y_1 y_2$ and $f^2 = y_2 y_3$ are both contained in a $j$-cycle, for $3 \leqslant j \leqslant q$. Since $V(C_m) \to S_1$, we deduce that they are also contained in a $j$-cycle for $q + 1 \leqslant j \leqslant m + q - 1$. In all, we see that both $f^1$ and $f^2$ belong to a $j$-cycle $C'_j$ for $j = 3, 4, \ldots, m + q - 1$, with $m + q - 1 \geqslant m + 2$, such that $V(C'_3) \subset V(C'_4) \subset \cdots \subset V(C'_{m+1})$. □